\documentclass[conference]{IEEEtran}
\IEEEoverridecommandlockouts
\usepackage[T1]{fontenc}
\usepackage{cite}
\usepackage[cmex10]{amsmath}
\usepackage{amsthm}
\usepackage{amssymb}
\usepackage{bbm}
\usepackage{mathtools}
\usepackage[hidelinks]{hyperref}
\usepackage{algorithmic}
\usepackage{textcomp}
\usepackage[usenames, dvipsnames]{color}
\usepackage[linesnumbered,ruled]{algorithm2e}
\usepackage{subcaption}
\usepackage{graphicx}  
\usepackage{url}
\usepackage[usestackEOL]{stackengine}
\usepackage{verbatim}
\usepackage{dsfont}

\usepackage {tikz}
\tikzset{>=stealth}
\usepackage{xcolor,colortbl}
\definecolor{lightred}{RGB}{255, 240, 240}
\definecolor{lightblue}{RGB}{230, 250, 255}
\definecolor{lightgreen}{RGB}{240, 255, 242}
\definecolor{myred}{RGB}{220, 0, 0}
\definecolor{myblue}{RGB}{0, 17, 173}
\definecolor{mygreen}{RGB}{2, 117, 0}

\newcommand{\mca}[1]{\mathcal{#1}}

\newcommand{\mbf}[1]{\mathbf{#1}}
\newcommand{\mbb}[1]{\mathbb{#1}}
\newcommand{\mrm}[1]{\mathrm{#1}}

\newcommand{\bsm}[1]{\boldsymbol{#1}}

\newcommand{\Tc}{T_{\mathrm{c}}}

\newcommand{\diag}[1]{\operatorname{diag}{#1}}

\newcommand{\op}[1]{\operatorname{#1}}

\DeclareMathOperator*{\maximize}{maximize}

\DeclareMathOperator*{\st}{subject\;to}

\ifCLASSINFOpdf
\else
\fi
\usepackage{amsmath}
\begin{document}
%
\title{One-Bit Massive MIMO Precoding for Frequency-Selective Fading Channels}

\author{\IEEEauthorblockN{Ly V. Nguyen\IEEEauthorrefmark{1}\IEEEauthorrefmark{2}, Lu Liu\IEEEauthorrefmark{1}, Nguyen Linh-Trung\IEEEauthorrefmark{2}, and A. Lee Swindlehurst\IEEEauthorrefmark{1}}
\IEEEauthorblockA{\IEEEauthorrefmark{1}Center for Pervasive Communications \& Computing, University of California, Irvine, CA, USA}
\IEEEauthorblockA{\IEEEauthorrefmark{2}AVITECH Institute, VNU University of Engineering and Technology, Hanoi, Vietnam}
Email: vanln1@uci.edu, liul22@uci.edu, linhtrung@vnu.edu.vn, swindle@uci.edu
\thanks{This work was supported by the National Science Foundation under grant CCF-2008724.}
}

\maketitle

\begin{abstract}
One-bit digital-to-analog converters (DACs) are a practical and promising solution for reducing cost and power consumption in massive multiple-input multiple-output (MIMO) systems. However, the one-bit precoding problem is NP-hard and even more challenging in frequency-selective fading channels compared to the flat-fading scenario. While block-wise processing (BWP) can effectively address the inter-symbol-interference (ISI) in frequency-selective fading channels, its computational complexity and processing delay can be too high for practical implementation. An alternative solution to alleviate the processing complexity and delay issues is symbol-wise processing (SWP) which sequentially designs the transmit signals. However, existing SWP work leaves unwanted interference for later signal designs. In this paper, we propose an SWP approach which can efficiently address the ISI even at the symbol rate. The idea is to design the transmit signal to not only be beneficial for its time slot, but also to provide constructive interference for subsequent symbols. We develop two active ISI processing methods that significantly outperform a conventional approach, one of which that even outperforms the BWP approach at low SNR.
\end{abstract}



%
\IEEEpeerreviewmaketitle

\section{Introduction}
\label{sec_introduction}
Massive multiple-input multiple-output (MIMO) technology is a key for 5G-and-beyond wireless networks due to the energy and spectral efficiency benefits that derive from employing very large antenna arrays at the base station (BS). However, cost and power consumption at the BS in massive MIMO systems can be prohibitively high when implemented with standard high-resolution radio-frequency hardware. The use of one-bit digital-to-analog converters (DACs) is an alternative solution that significantly reduces cost and power consumption in massive MIMO systems. Unfortunately, optimal one-bit massive MIMO precoding is an NP-hard problem because each antenna can only transmit a symbol in the set $\{\pm1 \pm1j\}$. This challenging but interesting problem has been studied intensively in the literature. However, the majority of exiting work consider flat-fading channels, e.g.,~\cite{Li2017Achievable,Jacobsson2017Quantized,Li2018Massive,Li2020Interference,Jedda2018Quantized_CEP,Li2021Onebit}. For frequency-selective fading channels, there has been some results reported in~\cite{ Hela2017Massive,Jedda2018Quantized,Nedelcu2018MAGIQ,Nedelcu2022low,Jacobsson2019Linear,Askerbeyli2019One-bit,Jacobsson2018Nonlinear}, but this work is primarily focused on orthogonal frequency division multiplexing (OFDM).

In this paper, we study the problem of one-bit massive MIMO precoding for frequency-selective fading channels. This problem is more challenging compared to flat-fading channels due to inter-symbol-interference (ISI), where symbols transmitted in one time slot affect the received signal at not only that time slot but those in the future. This line of research can be categorized into two groups: symbol-wise processing (SWP) and block-wise processing (BWP). In SWP, the transmit signals in different time slots of a coherence block are designed sequentially and separately~\cite{Hela2017Massive}, while in BWP they are jointly optimized~\cite{ Hela2017Massive,Jedda2018Quantized,Nedelcu2018MAGIQ,Nedelcu2022low,Jacobsson2019Linear,Askerbeyli2019One-bit,Jacobsson2018Nonlinear}. The main benefit of BWP is that ISI can be effectively addressed thanks to the joint optimization over the entire block. However, such approaches suffer from high computational complexity and long processing delay because the design of all the transmit signals in the block must be done concurrently before the signal in the first time slot can be transmitted. On the other hand, SWP can alleviate both the complexity and processing delay associated with BWP since it designs the transmit signals independently from one time slot to the next. For SWP, once the transmit signal in a given time slot is designed, it can be transmitted without waiting for the design of future signals. However, SWP is inferior to BWP in terms of performance since it cannot fully address the ISI.

To the best of our knowledge,~\cite{Hela2017Massive} is the only work in the literature of one-bit massive MIMO precoding for frequency-selective fading channels that has considered the SWP approach. However, the SWP algorithm in~\cite{Hela2017Massive} does not take into account the effects of the transmitted signals on later time slots. Motivated by this observation, in this paper we propose an SWP approach that can efficiently address the ISI effect even at the symbol rate. The idea is to design the transmit signal to not only be beneficial for its time slot, but also to provide constructive interference for subsequent symbols. We propose two SWP methods based on the maximum-safety margin optimization metric, one of which outperforms the other at low signal-to-noise ratios (SNRs) and vice versa at high SNRs. Simulation results also show that the bit-error-rate (BER) of the proposed methods are significantly lower than that of the conventional SWP method in~\cite{Hela2017Massive} and one of the proposed methods even outperforms the corresponding BWP approach at low SNRs.

\textit{Notation:} Upper-case and lower-case boldface letters denote matrices and column vectors, respectively. $|\cdot|$ denotes the absolute value of a number and $[\cdot]^T$ denotes the transpose. The notation $\Re\{\cdot\}$ and $\Im\{\cdot\}$ respectively denotes the real and imaginary parts of the complex argument. If $\Re\{\cdot\}$ and $\Im\{\cdot\}$ are applied to a matrix or vector, they are applied separately to every element of that matrix or vector. $\mathbb{R}$ and $\mathbb{C}$ denote the set of real and complex numbers, respectively, and $j$ is the unit imaginary number satisfying $j^2=-1$.

\section{System Model and Problem Formulation}
\label{sec_system_model_and_problem_formulation}
\subsection{System Model}
We consider a downlink massive MIMO system with an $N$-antenna base station serving $K$ single-antenna users, where it is assumed that $N \geq K$.  Let $\mbf{H}_\ell \in \mbb{C}^{K\times N}$ denote the $\ell^\text{th}$ channel tap, $\ell \in \mca{L} = \{0,1,\ldots,L-1\}$, where $L$ is the number of channel taps. We assume perfect channel state information (CSI) and focus on the precoding problem. Let $\mbf{x}_t$ denote the transmit signal vector at time slot $t$. We assume that the base station employs two $1$-bit DACs, one for the in-phase and the other for the quadrature signal. Hence, the signal $x_{t,n}$ transmitted by the $n^{\text{th}}$ antenna is confined to the discrete set $\mathcal{X} = \{\pm 1 \pm1j\}$. Let $\mbf{y}_t \in \mbb{C}^K$ be the signal vector received by the users, which is given as
\begin{equation}
    \mbf{y}_t = \sqrt{\frac{\rho}{2N}} \sum_{\ell=0}^{L-1} \mbf{H}_\ell\mbf{x}_{t-\ell} + \mbf{n}_t,
    \label{eq_freq_selec_sys_model}
\end{equation}
where $\mbf{n}_t\sim\mca{CN}(0,\sigma^2\mbf{I}_K)$ is the noise vector, $t = 1,\ldots,\Tc$, where $\Tc$ is the length of the coherence block, and the normalization by $2N$ leads to the interpretation of $\rho$ as the total transmit power.

\subsection{Problem Formulation}
\label{subsec_problem_formulation}
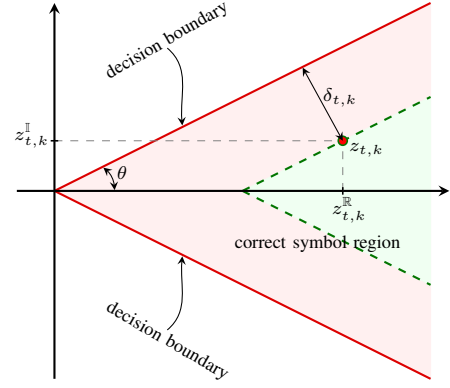
\begin{figure}
    \centering
    \begin{tikzpicture}
    \draw [lightred, fill=lightred] (0,0) -- (5,2.5) -- (5,-2.5) -- (0,0);

    \draw [lightgreen, fill=lightgreen] (2.5,0) -- (5,1.25) -- (5,-1.25) -- (2.5,0);
    
    \draw [thin, <->] (0.65,0.31) to [out=-40, in=95] (0.8,0.01);
    \node at (0.9,0.25) {{\scriptsize $\theta$}};
    
    \draw [thin, ->] (1.5,1.9) to [out=-40, in=95] (1.7,0.87);
    \node [rotate = 30] at (1.5,2) {{\scriptsize decision boundary}};
    
    \draw [thin, ->] (1.5,-1.85) to [out=40, in=-95] (1.7,-0.87);
    \node [rotate = -30] at (1.5,-2) {{\scriptsize decision boundary}};
    
    \draw [thick,-,myred] (0,0) to (5,2.5);
    \draw [thick,-,myred] (0,0) to (5,-2.5);
    
    \draw [thick,dashed,-,mygreen] (2.5,0) to (5,1.25);
    \draw [thick,dashed,-,mygreen] (2.5,0) to (5,-1.25);
    
    \draw [thick,->] (-0.5,0) to (5.25,0);
    \draw [thick, ->] (0,-2.5) to (0,2.5);
    
    \draw [help lines, dashed, -] (0,0.667) to (3.833,0.667);
    \draw [help lines, dashed, -] (3.833,0) to (3.833,0.667);
    \draw [mygreen, fill = red, semithick] (3.833,0.667) circle [radius=0.06];
    \node at (4.15,0.55) {{\scriptsize ${z}_{t,k}$}};
    
    \draw [thin,<->] (3.833,0.667) to (3.3,1.64);
    \node at (3.8,1.25) {{\scriptsize $\delta_{t,k}$}};
    
    \draw [semithick,-] (3.833,-0.05) to (3.833,0.05);
    \draw [semithick,-] (-0.05,0.667) to (0.05,0.667);
    \node at (3.92,-0.25) {{\scriptsize  $z_{t,k}^{\mbb{R}}$}};
    \node at (-.3,0.74) {{\scriptsize  $z_{t,k}^{\mbb{I}}$}};

    \node at (3.5,-0.7) {{\scriptsize  correct symbol region}};
    
    \end{tikzpicture}
    \caption{Illustration of the safety margin for user $k$ at time slot $t$. The correct symbol region includes the pink and green areas.}
    \label{fig_safety_margine_illustration}
\end{figure}

Let $\mbf{s}_t \in \mca{C}^K$ denote the symbols we intend the users to detect at time slot $t$. We consider $D$-PSK signaling, i.e., $s_{t,k} \in \exp{(j\pi\frac{2d_k+1}{D})}$ where $d_k \in \{0,\ldots,D-1\}$. The rotated noiseless received signal vector is given as
\begin{align}
    \mbf{{z}}_t = \gamma \diag(\mbf{{s}}^*_t)\sum_{\ell=0}^{L-1}\mbf{H}_\ell\mbf{x}_{t-\ell}
    \label{eq_intermediate_signal_zt}
\end{align}
where $\gamma = \sqrt{\rho/(2N)}$. The safety margin~\cite{Jedda2018Quantized} of user $k$ at time slot $t$ is illustrated in Fig.~\ref{fig_safety_margine_illustration} and is given by 
\begin{align}
    \delta_{t,k} = z^\mbb{R}_{t,k}  \sin(\theta) - |z^\mbb{I}_{t,k} |\cos(\theta),
\end{align}
where $z^\mbb{R}_{t,k}$ and $z^\mbb{I}_{t,k}$ denote the real and imaginary parts of $z_{t,k}$, respectively, and $\theta = \pi/D$. It is clear that the farther $z_{t,k}$ is from the symbol decision boundaries, the more likely that the received signal $y_{t,k}$ will be correctly detected, i.e., the more robust it will be against the effects of noise and interference. Therefore, we want to increase the safety margins of the users as much as possible.

A common design approach is to maximize the minimum safety margin $\min \delta_{t,k}$ over the users and over the entire coherence block. However, this approach requires block-wise processing of all the transmit signal vectors $\{\mbf{x}_1,\ldots,\mbf{x}_{\Tc}\}$. Such as a block-wise design can lead to excessive computational complexity and processing delay since the signal in the first time slot $\mbf{x}_1$ cannot be transmitted until the entire block design is completed. For example, BWP based on linear programming scales polynomially with the block size $\Tc$ while SWP scales only linearly with $\Tc$~\cite{Hela2017Massive}. In this paper, we focus on the SWP design perspective and propose two methods that can effectively address the ISI effect.

\section{Passive ISI Processing}
This section presents the SWP design method in~\cite{Hela2017Massive}, which is referred to as \textit{passive} ISI processing. The received signal vector at time slot $t$ can be decomposed as follows:
\begin{equation}
    \mbf{y}_t = \gamma \mbf{H}_0\mbf{x}_t + \gamma \underbrace{\sum_{\ell=1}^{L-1}\mbf{H}_{\ell}\mbf{x}_{t-\ell}}_{\bsm{\eta}_t} + \mbf{n}_t \; ,
    \label{eq:rx_signal_time_t}
\end{equation}
where the term $\bsm{\eta}_t$ represents the ISI due to the delayed channel taps. The rotated noiseless received signal vector can be then written in the following form:
\begin{align}
    \mbf{{z}}_t &= \diag(\mbf{{s}}^*_t)\bigg(\gamma\mbf{{H}}_0\mbf{{x}}_t + \gamma\sum_{\ell=1}^{L-1}\mbf{H}_{\ell}\mbf{x}_{t-\ell}\bigg) \\
    &= \mbf{{W}}_t \mbf{x}_t + \mbf{u}_t \; ,
    \label{eq_intermediate_signal_P_t}
\end{align}
where $\mbf{W}_t=\gamma\diag(\mbf{{s}}^*_t)\mbf{H}_0$ reflects the effect of the current channel tap $\mbf{H}_0$ and $\mbf{u}_t = \gamma\diag(\mbf{s}^*_t)\sum_{\ell=1}^{L-1}\mbf{H}_{\ell}\mbf{x}_{t-\ell}$ accounts for the ISI due to the delayed channel taps.

At a time slot $t$, the SWP design optimizes the transmit signal vector $\mbf{x}_t$ to maximize the minimum safety margin of this time slot~\cite{Hela2017Massive}, which can be written as
\begin{equation}
\begin{aligned}
& \maximize_{\mbf{x}_t,\;\delta^{\mrm{min}}}
& & \delta^{\mrm{min}}\\
& \operatorname{subject\ to}
& & \delta_{t,k} \geq \delta^{\mrm{min}} \;\; \forall k \in \mca{K},\\
& & & \mathbf{x}_t \in \{\pm 1\}^{2N}.
\end{aligned}
\label{eq:passive_SWP_2}
\end{equation}
The constraint $\delta_{t,k} \geq \delta^{\mrm{min}} \; \forall k\in\mca{K}$ can be written in the matrix form $\mbf{Q}_t\bsm{\nu}_t \leq \mbf{c}_t$, where $\bsm{\nu}_t = [\Re\{\mbf{x}^T_t\},\Im\{\mbf{x}^T_t\}, \delta^{\mrm{min}}]^T$ is the vector variable to be optimized, $\mbf{c}_t$ is a vector accounting for the ISI and is given as
\begin{align}
    \mbf{c}_t = \begin{bmatrix}
    \tan(\theta)\Re\{\mbf{u}_t\} - \Im\{\mbf{u}_t\} \\
    \tan(\theta)\Re\{\mbf{u}_t\} + \Im\{\mbf{u}_t\}
    \end{bmatrix},
\end{align}
and
\begin{equation}
    \mbf{Q}_t =  \begin{bmatrix}
\mbf{B}_t - \tan(\theta)\mbf{A}_t & \frac{1}{\cos (\theta)}\mbf{1}_K\\
-\mbf{B}_t - \tan(\theta)\mbf{A}_t & \frac{1}{\cos (\theta)}\mbf{1}_K
\end{bmatrix} \; ,
\end{equation}
where $\mbf{A}_t = [ \Re \{\mbf{{W}}_t\}, -\Im \{\mbf{{W}}_t\}]$ and $\mbf{B}_t  = [\Im \{\mbf{{W}}_t \}, \Re \{\mbf{{W}}_t\}]$.
In~\cite{Hela2017Massive}, the constraints $x_{t,k} \in \{\pm1\}$ are relaxed to $-1\leq x_{t,k}\leq 1$ to obtain the following convex linear programming problem:
\begin{equation}
\begin{aligned}
& \maximize_{\bsm{\nu}_t}
& & [\mbf{0}^T_{2N},\;1]^T\bsm{\nu}_t\\
& \st
& &\mbf{Q}_t\bsm{\nu}_t \leq 
\mbf{c}_t\\
& & & -\mbf{1}_{2N} \leq \begin{bmatrix}
\Re\{\mbf{x}_t\} \\ \Im\{\mbf{x}_t\}
\end{bmatrix} \leq \mbf{1}_{2N} \; .
\end{aligned}
\label{eq:passive_SWP_3}
\end{equation}
If we let $\bsm{{\nu}}^\star_t$ be the solution of~\eqref{eq:passive_SWP_3}, the transmit signal $\mbf{x}_t$ is obtained as $x_{t,n} = \op{sign}({\nu}^\star_{t,n})$ for $n = 1,\ldots,2N$.

\begin{figure}
    \centering
    \includegraphics[width=0.9\linewidth]{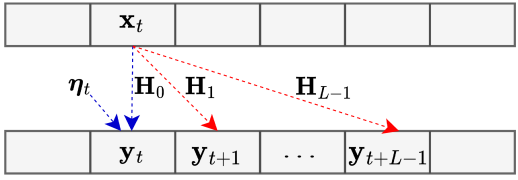}
    \caption{The design of $\mbf{x}_t$ in~\cite{Hela2017Massive} only takes into account the ISI term $\bsm{\eta}_t$ and the effect of $\mbf{x}_t$ on the received signal at time $t$ (blue arrows), and ignores the effect of $\mbf{x}_t$ on the future time slots (red arrows).}
    \label{fig:ISI_illustration}
\end{figure}

\textit{Discussion:} In the above SWP approach, the ISI term $\bsm{\eta}_t$ from the past transmit signals and the effect of $\mbf{x}_t$ on time slot $t$ are taken into account when designing the signal $\mbf{x}_t$. However, this method ignores the effect of $\mbf{x}_t$ on the future (delayed) time slots $t+1,\ldots,t+L-1$ as illustrated in Fig.~\ref{fig:ISI_illustration}, and therefore unintentionally induces unwanted interference for the design of the future signals $\mbf{x}_{t+1},\ldots,\mbf{x}_{t+L-1}$. In other words, the design of $\mbf{x}_t$ has to \textit{passively} cope with the ISI term $\bsm{\eta}_t$ which is unwanted interference from the design of $\mbf{x}_{t-1},\ldots,\mbf{x}_{t-L+1}$. Motivated by this observation, in the following section, we propose an SWP approach that takes into account $\bsm{\eta}_t$ and the effect of $\mbf{x}_t$ on all time slots from $t$ to $t + L -1$. In this way, our proposed approach will \textit{actively} provide constructive interference for the future signal designs.

\section{Proposed Active ISI Processing}
Here, we propose an SWP approach that takes into account the interference of the past time slots while at the same time providing constructive interference for those in the future. Since the signal $\mbf{x}_t$ affects the $L$ time slots $t,\ldots,t+L-1$, our idea is to take into account the safety margins at these time slots when designing $\mbf{x}_t$. This is unlike the method in~\cite{Hela2017Massive} which considers the safety margins only at time slot $t$ when designing $\mbf{x}_t$. In the following, we propose two relevant optimization methods; one maximizes the minimum safety margin over all the users and time slots $t,\ldots,t+L-1$, while the other maximizes the sum of the minimum safety margins obtained over the time slots $t,\ldots,t+L-1$.
\subsection{Method 1: Maximizing the Minimum Safety Margin}
This method aims to maximize the minimum safety margin of all $K$ users over $L$ time slots $t,\ldots,t+L-1$ as follows:
\begin{equation}
\begin{aligned}
& \maximize_{\mbf{x}_t,\;\delta^{\mrm{min}}}
& & \delta^{\mrm{min}}\\
& \st
& & \delta_{t+\ell,k} \geq \delta^{\mrm{min}} \;\; \forall \ell\in\mca{L}, \;k \in \mca{K}\\
& & & \mathbf{x}_t \in \{\pm 1\}^{2N}.
\end{aligned}
\end{equation}
This optimization problem can also be relaxed and written as a linear programming problem:
\begin{equation}
\begin{aligned}
& \maximize_{\bsm{\nu}_t}
& & [\mbf{0}^T_{2N},\;1]^T\bsm{\nu}_t\\
& \st
& &\mbf{Q}_{t+\ell} \bsm{\nu}_t \leq \mbf{c}_{t+\ell} \;\; \forall \ell\in\mca{L}\\
& & & -\mbf{1}_{2N} \leq \begin{bmatrix}
\Re\{\mbf{x}_t\} \\ \Im\{\mbf{x}_t\}
\end{bmatrix} \leq \mbf{1}_{2N}.
\end{aligned}
\label{eq:proposed_SWP1}
\end{equation}
Note that the definition of $\mbf{Q}_{t+\ell}$ requires $\mbf{{A}}_{t+\ell}$ and $\mbf{{B}}_{t+\ell}$, which are given by
$\mbf{A}_{t+\ell} = \begin{bmatrix} \Re \{\mbf{{W}}_{t+\ell}\} & -\Im \{\mbf{{W}}_{t+\ell}\} \end{bmatrix}$ and $\mbf{B}_{t+\ell} = \begin{bmatrix} \Im \{\mbf{{W}}_{t+\ell} \} & \Re \{\mbf{{W}}_{t+\ell}\} \end{bmatrix}$,
where $\mbf{{W}}_{t+\ell} = \diag(\mbf{s}^*_{t+\ell})\mbf{H}_{\ell}$. The definition of $\mbf{c}_{t+\ell}$ requires $\mbf{u}_{t+\ell}$, which is given by 
$\mbf{u}_{t+\ell} = \diag(\mbf{{s}}^*_{t+\ell})\sum_{\ell'=\ell+1}^{L-1}\mbf{H}_{\ell'}\mbf{x}_{t+\ell-\ell'}.$
It should be noted that the signals $\mbf{x}_{t+1}, \ldots, \mbf{x}_{t+L-1}$ have not been designed yet, and therefore the safety margins at time slots $t+1,\ldots,t+L-1$ are computed using only the previously designed signals $\mbf{x}_{t-1}, \ldots, \mbf{x}_{t-L+2}$. This explains why the index $\ell'$ in the computation of $\mbf{u}_{t+\ell}$ starts from $\ell+1$ instead of $1$. Finally, we take the sign of the first $2N$ elements of the solution of~\eqref{eq:proposed_SWP1} to obtain the transmit signal $\mbf{x}_t$.

\subsection{Method 2: Maximizing the Sum of Minimum Safety Margins}
This method aims to maximize the sum of the per-time slot minimum safety margins, as follows:
\begin{equation}
\begin{aligned}
& \maximize_{\mbf{x}_t,\;\delta_{\ell}^{\mrm{min}}}
& & \sum_{\ell=0}^{L-1}\delta_{\ell}^{\mrm{min}}\\
& \st
& & \delta_{t+\ell,k} \geq \delta_{\ell}^{\mrm{min}} \;\; \forall \ell\in\mca{L}, \;k \in \mca{K}\\
& & & \mathbf{x}_t \in \{\pm 1\}^{2N} \; .
\end{aligned}
\end{equation}
This problem can also be relaxed and written as a linear programming problem:
\begin{equation}
\begin{aligned}
& \maximize_{\bsm{\upsilon}_t}
& & [\mbf{0}^T_{2N},\;\mbf{1}^T_L]^T\bsm{\upsilon}_t\\
& \st
& &\mbf{G}_{t+\ell}\bsm{\upsilon}_t \leq \mbf{c}_{t+\ell} \;\; \forall \ell\in\mca{L}\\
& & & -\mbf{1}_{2N} \leq \begin{bmatrix}
\Re\{\mbf{x}_t\} \\ \Im\{\mbf{x}_t\}
\end{bmatrix} \leq \mbf{1}_{2N}.
\end{aligned}
\label{eq:proposed_SWP2}
\end{equation}
Here, $\bsm{\upsilon}_t = [
\Re\{\mbf{x}^T_t\}, \Im\{\mbf{x}^T_t\}, \delta^{\mrm{min}}_0, \cdots ,\delta^{\mrm{min}}_{L-1}]^T$ and
\begin{equation}
    \mbf{G}_{t+\ell} =  \begin{bmatrix}
\mbf{B}_{t+\ell} - \tan(\theta)\mbf{A}_{t+\ell} & \frac{1}{\cos (\theta)}\mbf{E}_{\ell+1}\\
-\mbf{B}_{t+\ell} - \tan(\theta)\mbf{A}_{t+\ell} & \frac{1}{\cos (\theta)}\mbf{E}_{\ell+1}
\end{bmatrix} \; ,
\end{equation}
where $\mbf{E}_{\ell+1}$ is a real-valued matrix of size $K\times L$ whose $(\ell+1)^{\text{th}}$ column is a vector of all ones and whose other columns are all zeros. Similarly, we take the sign of the first $2N$ elements of the solution of~\eqref{eq:proposed_SWP2} to obtain the transmit signal $\mbf{x}_t$.

\section{Numerical Results}
\label{sec_numerical_results}
This section provides numerical results to show the superiority of the proposed methods. We set $K = 4$, $N = 64$, $\Tc = 256$, and $D = 8$ (i.e., 8-PSK signaling). Each channel element is generated as a $\mca{CN}(0,1/L)$ random variable and the SNR is defined as $\rho/\sigma^2$.
\begin{figure}
    \centering
    \includegraphics[scale=0.55]{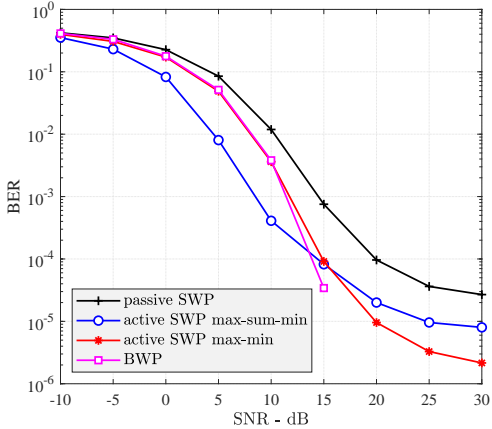}
    \caption{BER performance comparison with $L = 3$.}
    \label{fig:BER}
\end{figure}

In Fig.~\ref{fig:BER}, we compare the proposed SWP methods 1 and 2 (referred to as `max-min' and `max-sum-min', respectively) with the conventional SWP method (referred to as `passive SWP') and also the BWP method in~\cite{Hela2017Massive}. It can be seen that the proposed methods significantly outperform the conventional passive SWP method, since the active SWP methods create constructive interference for the transmit signal design in future symbol periods to exploit. It is also interesting to note that the max-sum-min method gives the best performance at low SNRs and even outperforms the BWP method which jointly designs the entire coherence block of 256 time slots. 

\begin{figure}
    \centering
    \includegraphics[scale=0.55]{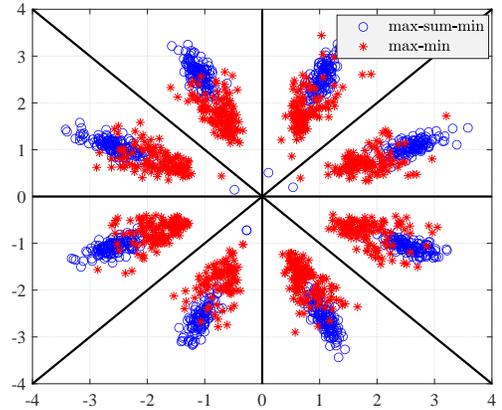}
    \caption{Noiseless received signals for the proposed methods.}
    \label{fig:rx_signal_constel}
\end{figure}

At high SNRs, the max-min method gives lower BERs compared to the max-sum-min. To explain this, we provide a sample plot of the noiseless received signals for the max-min and max-sum-min methods in Fig.~\ref{fig:rx_signal_constel}. It is observed that while the max-sum-min method moves the majority of signals far from the decision boundaries, the max-min method pushes the worst signal sample away from the boundaries and therefore the majority of signals are pulled closer to the decision thresholds as compared to the max-sum-min method. This explains why at low SNRs, when the noise is strong, the max-sum-min approach gives better performance. However, the drawback of the max-sum-min method is that it focuses on the strongest signals and therefore may leave some received signals very near the origin, as seen in the figure. Such signals are obviously more susceptible to a noise-induced detection error. 

\begin{figure}
    \centering
    \includegraphics[scale=0.6]{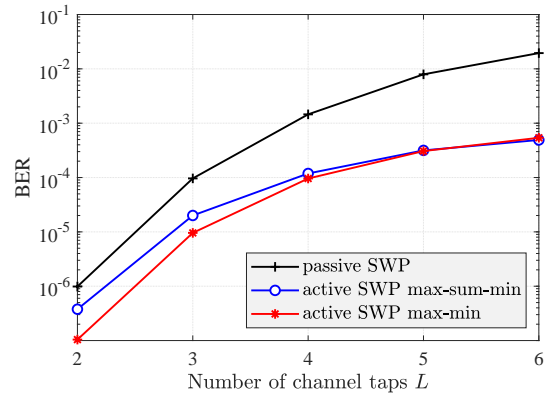}
    \caption{BER performance comparison for different values of $L$ at 20-dB.}
    \label{fig:BER_versus_L}
\end{figure}
In Fig.~\ref{fig:BER_versus_L}, we compare the proposed active SWP methods with the conventional passive SWP method for different numbers of channel taps $L$. It can be seen that as $L$ increases, the improvement between the proposed active methods and the conventional passive method also increases. This is due to the fact that a channel with a longer delay spread will result in more ISI, which significantly degrades the performance of the passive method since it ignores the future effect of a the design in a given time slot on future time slots. On the other hand, the proposed active methods better account for this effect by actively providing constructive interference that can be exploited in the design of future transmitted signals.

\section{Conclusion}
\label{sec_conclusion}
In this paper, we have proposed an SWP approach that not only takes into account interference from past signals on the current time slot, but that also generates constructive interference that can be exploited by future signal designs. We proposed two active ISI precoders, one based on maximizing the minimum safety margin for all users, and the other on maximizing the sum of the minimum safety margins over the delay spread. These two methods effectively address the ISI effect even at the symbol processing rate and significantly outperform a conventional SWP method. One of the proposed SWP methods can even yield better performance compared to its BWP counterpart at low SNRs.

  \newpage

\bibliographystyle{IEEEtran}
\bibliography{ref}

\begin{thebibliography}{10}
\providecommand{\url}[1]{#1}
\csname url@samestyle\endcsname
\providecommand{\newblock}{\relax}
\providecommand{\bibinfo}[2]{#2}
\providecommand{\BIBentrySTDinterwordspacing}{\spaceskip=0pt\relax}
\providecommand{\BIBentryALTinterwordstretchfactor}{4}
\providecommand{\BIBentryALTinterwordspacing}{\spaceskip=\fontdimen2\font plus
\BIBentryALTinterwordstretchfactor\fontdimen3\font minus
  \fontdimen4\font\relax}
\providecommand{\BIBforeignlanguage}[2]{{%
\expandafter\ifx\csname l@#1\endcsname\relax
\typeout{** WARNING: IEEEtran.bst: No hyphenation pattern has been}%
\typeout{** loaded for the language `#1'. Using the pattern for}%
\typeout{** the default language instead.}%
\else
\language=\csname l@#1\endcsname
\fi
#2}}
\providecommand{\BIBdecl}{\relax}
\BIBdecl

\bibitem{Li2017Achievable}
Y.~Li, C.~Tao, A.~Lee~Swindlehurst, A.~Mezghani, and L.~Liu, ``Downlink
  achievable rate analysis in massive {MIMO} systems with one-bit {DAC}s,''
  \emph{IEEE Commun. Letters}, vol.~21, no.~7, pp. 1669--1672, July 2017.

\bibitem{Jacobsson2017Quantized}
S.~Jacobsson, G.~Durisi, M.~Coldrey, T.~Goldstein, and C.~Studer, ``Quantized
  precoding for massive {MU-MIMO},'' \emph{IEEE Trans. Commun.}, vol.~65,
  no.~11, pp. 4670--4684, Nov. 2017.

\bibitem{Li2018Massive}
A.~Li, C.~Masouros, F.~Liu, and A.~L. Swindlehurst, ``Massive {MIMO} 1-bit
  {DAC} transmission: {A} low-complexity symbol scaling approach,'' \emph{IEEE
  Trans. Wireless Commun.}, vol.~17, no.~11, pp. 7559--7575, Nov. 2018.

\bibitem{Li2020Interference}
A.~Li, F.~Liu, C.~Masouros, Y.~Li, and B.~Vucetic, ``Interference exploitation
  1-bit massive {MIMO} precoding: {A} partial branch-and-bound solution with
  near-optimal performance,'' \emph{IEEE Trans. Wireless Commun.}, vol.~19,
  no.~5, pp. 3474--3489, May 2020.

\bibitem{Jedda2018Quantized_CEP}
H.~Jedda, A.~Mezghani, A.~L. Swindlehurst, and J.~A. Nossek, ``Quantized
  constant envelope precoding with {PSK} and {QAM} signaling,'' \emph{IEEE
  Trans. Wireless Commun.}, vol.~17, no.~12, pp. 8022--8034, Dec. 2018.

\bibitem{Li2021Onebit}
A.~Li, C.~Masouros, A.~L. Swindlehurst, and W.~Yu, ``1-bit massive {MIMO}
  transmission: {E}mbracing interference with symbol-level precoding,''
  \emph{IEEE Commun. Magazine}, vol.~59, no.~5, pp. 121--127, May 2021.

\bibitem{Hela2017Massive}
H.~Jedda, A.~Mezghani, J.~A. Nossek, and A.~L. Swindlehurst, ``Massive {MIMO}
  downlink 1-bit precoding for frequency selective channels,'' in \emph{Proc.
  IEEE Int. Workshop Computational Advances in Multi-Sensor Adaptive Process.},
  Curacao, Dec. 2017.

\bibitem{Jedda2018Quantized}
H.~Jedda and J.~A. Nossek, ``Quantized constant envelope precoding for
  frequency selective channels,'' in \emph{Proc. IEEE Statistical Signal
  Process. Workshop}, Freiburg im Breisgau, Germany, June 2018, pp. 213--217.

\bibitem{Nedelcu2018MAGIQ}
A.~Nedelcu, F.~Steiner, M.~Staudacher, G.~Kramer, W.~Zirwas, R.~S. Ganesan,
  P.~Baracca, and S.~Wesemann, ``Quantized precoding for multi-antenna downlink
  channels with {MAGIQ},'' in \emph{Proc. Int. ITG Workshop on Smart Antennas},
  Bochum, Germany, Mar. 2018.

\bibitem{Nedelcu2022low}
A.~S. Nedelcu, F.~Steiner, and G.~Kramer, ``Low-resolution precoding for
  multi-antenna downlink channels and {OFDM},'' \emph{Entropy}, vol.~24, no.~4,
  p. 504, 2022.

\bibitem{Jacobsson2019Linear}
S.~Jacobsson, G.~Durisi, M.~Coldrey, and C.~Studer, ``Linear precoding with
  low-resolution {DAC}s for massive {MU-MIMO-OFDM} downlink,'' \emph{IEEE
  Trans. Wireless Commun.}, vol.~18, no.~3, pp. 1595--1609, Mar. 2019.

\bibitem{Askerbeyli2019One-bit}
F.~Askerbeyli, H.~Jedda, and J.~A. Nossek, ``1-bit precoding in massive
  {MU-MISO-OFDM} downlink with linear programming,'' in \emph{Proc. Int. ITG
  Workshop on Smart Antennas}, Vienna, Austria, Apr. 2019.

\bibitem{Jacobsson2018Nonlinear}
S.~Jacobsson, O.~Casta{\~{n}}eda, C.~Jeon, G.~Durisi, and C.~Studer,
  ``Nonlinear precoding for phase-quantized constant-envelope massive
  {MU-MIMO-OFDM},'' in \emph{Proc. Int. Conf. Telecommun.}, Saint-Malo, France,
  June 2018, pp. 367--372.

\end{thebibliography}


%









\end{document}